\documentclass[%
 reprint,
 amsmath,amssymb,
 aps,
]{revtex4-1}
\usepackage{float}
\usepackage{amsthm}
\usepackage{graphicx}
\usepackage{dcolumn}
\usepackage{bm}

\begin{document}

\preprint{APS/123-QCC}

\title{Side information -driven quantum composite control for protecting a qubit}

\author{Ya Cao$^{1}$}
 \email{caoshinee@126.com}
\author{Fei Gao$^{1}$}
\author{DanDan Li$^{1}$}
\author{QiaoYan Wen$^{1}$}
 \email{}

\affiliation{%
 $^{1}$State Key Laboratory of Networking and Switching Technology, Beijing University of Posts and Telecommunications, Beijing, 100876, China\\
}%

\date{\today}

\begin{abstract}
We study protection of a qubit that transfer through a decoherence noise by quantum control technique. In this work, we assume that the communication participants have some side information about the qubit. Our aim is to take fully advantage of given side information to design a quantum control scheme that makes the output state close to the input state as better as possible. Especially, in this paper, we consider the quantum composite control (QCC) structure, which has shown to capture a better balance between fidelity and success probability, i.e., two essential measures for protection. Comparison has been made between our scheme and previous ones. We show there's advantages in two aspects. On one hand, when compared with the scheme that has taking fully consideration of classical side information but not quantum one, our scheme achieves higher fidelity. Inversely, when the scheme with quantum side information but not classical one is considered, our scheme also has nontrivial improvement.

\begin{description}
\item[PACS numbers]
03.67.Pp, 03.65.Ta, 02.30.Yy
\end{description}
\end{abstract}

\pacs{Valid PACS appear here}
\maketitle


\section{\label{sec:level1}Introduction\protect}
Quantum technologies have many advantages over their classical counterparts. For example, quantum key distribution (QKD) can achieve unconditional security. In these quantum techniques, quantum states play an irreplaceable role in bearing information for communication or computation. Unfortunately, there is no isolated system in the world. That means the state will be inevitably exposed to decoherence caused by interacton between environment and system, which will then make the system derivation from the expected state and thus disturbing quantum techniques from functioning in desired ways.

In classical realm, states can be easily protected against the noise by simply recording it before the noise and reconstruct it after the noise. In quantum world, it's not the case even when two nonorthogonal pure states are considered. Quantum control is one efficient technique to achieve this task. There are some fundamental features of quantum mechanics restricting the ability of quantum control. Firstly, Heisenberg's inequality limits the amount of information that measurement can obtain. This implies that two or more nonorthogonal stats cannot be perfectly discriminated. Secondly, there will be an undesired phenomenon, ``back-action'', when quantum measurements are applied. That is to say, once a measurement is taken, the system will be immediately disrupted and collapse to some unexpected state. These restrictions make us to design quantum control schemes delicately.

We would like to refer to recent works on quantum control for quantum state protection. In 2007, Branczyk $\emph{et al.}$ proposed a quantum feedback control (QFBC) to protect a qubit in either of two states, which are randomly prepared nonorthogonal states lying in the $XZ$ plane of the Bloch sphere, against the dephasing noise. This work demonstrated that weak measurement can balance the information gain and back-action and thus is beneficial for designing quantum control schemes. It has attracted much attention and has been experimentally implemented in 2010. Related works thereafter mainly considered in two aspects, protecting more states and defending more noises. These two items are cross presented and we will not introduce them in time line. For defending more noises, Xiao $\emph{et al.}$ reexamined the QFBC in all typical types of two-dimensional noise sources including bit-flip, amplitude-damping, dephasing and depolarization noises. QFBC can work but not too well in other noises than dephasing. Korotkov $\emph{et al.}$ presented that quantum measurement before noise and its reversal after noise can almost completely suppress the amplitude-damping noise. Inspired by this, Wang $\emph{et al.}$ proposed a quantum feedforward control (QFFC) scheme to protect two nonorthogonal states, which lies in spatial plane of the Bloch sphere, with equal a priori probability. This scheme can reach almost unit fidelity for all states and even heavy noises. However this happens at price of low success probability.

It should be emphasised that success probability is also an important measure to estimate the performance of a control scheme, not only fidelity. Comparing to another relevant work for quantum state protection, quantum error correction, quantum control has one advantage that it doesn't need redundant degree of freedom to encode the state of system into that is immune to the noise or can be corrected afterwards. More redundant degrees mean more resource consumption, so as low success probability. To solve the ``high fidelity with low success probability'' problem in QFFC scheme, our last work presented a quantum composite control (QCC) scheme by deliberately combine QFFC and QFBC procedures. This scheme possess the advantage of QFFC in fidelity, and generously improves the success probability.

In this paper, we reinvestigate the QCC technique to protect a qubit in more general states, $\{s_{\pm},|\psi_{\pm}\rangle\}$, where    $|\psi_{\pm}\rangle=\cos\frac{\theta}{2}|+\rangle +e^{i\phi}\sin\frac{\theta}{2}|-\rangle$ with $\{s_+,s_-\}$ an arbitrary priori probability.

We introduce different preweak and postweak measurements, feedforward and feedback operations for the quantum composite control. We improve the performance in two aspects, definite and indefinite protection schemes. Both schemes are considered with optimal parameters. Results show that this generalized scheme has advantages not only in inequal a priori probability case, but also in equal one in the sense that, with fixed fidelity it can obtain the highest success probability, and vice verse.
This proposal can also be used to protect a qubit in a mixed state $\rho=s_+|\psi_+\rangle\langle+|+s_-|\psi_-\rangle\langle-|$.

This paper is structured as below. Sec. \uppercase\expandafter{\romannumeral2} introduces the generalized quantum composite control scheme and two importance measures estimating the performance of protection. Sec. \uppercase\expandafter{\romannumeral3} analyzes the performance of this scheme in two directions, definite protection and indefinite protection. Sec. \uppercase\expandafter{\romannumeral3} concludes this paper.

\section{\label{sec:level2} Quantum composite control scheme for arbitrary priori probability scenario}

In what follows, we present the scheme for protecting a qubit in either of two nonorthogonal states with an arbitrary priori probability. The process is illustrated in Fig. 1.

\subsection{The Scheme}

Suppose the qubit sending from Alice to Bob through a decoherence channel is in either of two nonorthogonal states with an arbitrary priori probability,
\begin{equation}
    \{s_{\pm},|\psi_{\pm}\rangle\}.
\end{equation}
where ${|{\psi }_{\pm }}\rangle=\cos \frac{\theta }{2}|+\rangle\pm \ {{e}^{i\phi }}\sin \frac{\theta }{2}|-\rangle$ and $s_\pm$ are corresponding  priori probabilities. It holds that $s_+ + s_-=1$. Especially, when $s_+=s_-=1/2$, It will be the case considered in previous papers [FBC,FFC,QCC].

Notice that if a control scheme can protect a qubit in such state, it can also protect it in the mixed state,
\begin{equation}
    \rho=s_+|\psi_+\rangle\langle\psi_+|+s_-|\psi_-\rangle\langle\psi_-|.
\end{equation}

The decoherence here we consider is the amplitude damping (AD) noise, which pollutes an arbitary state $\rho$ into a mixed state
 \begin{equation}
\varepsilon(\rho)=E_{1}\rho E_{1}^{\dagger}+E_{2}\rho E_{2}^{\dagger},
\end{equation}
where
\begin{equation}
E_{1}=\left( \begin{array}{cc} 1&0\\ 0&\sqrt{1-r} \end{array} \right),  \ \
E_{2}=\left( \begin{array}{cc} 0&\sqrt{r}\\ 0&0 \end{array} \right),
\end{equation}
with a parameter $r$ representing the strength of noise.

A quantum control scheme aims to output state as close as possible to the primary one with just measurements and unitary operators, and no redundant systems. To protect the qubit state of interest, we propose a quantum composite control scheme as illustrated in FIG. 1, by taking full advantage of all known information. The quantum composite control scheme contains three procedures: a feedforward control(FFC) procedure, a reversal procedure and a feedback control(FBC) procedure.


\begin{figure*}[htbp]
\centering
  \includegraphics[width=0.8\paperwidth]{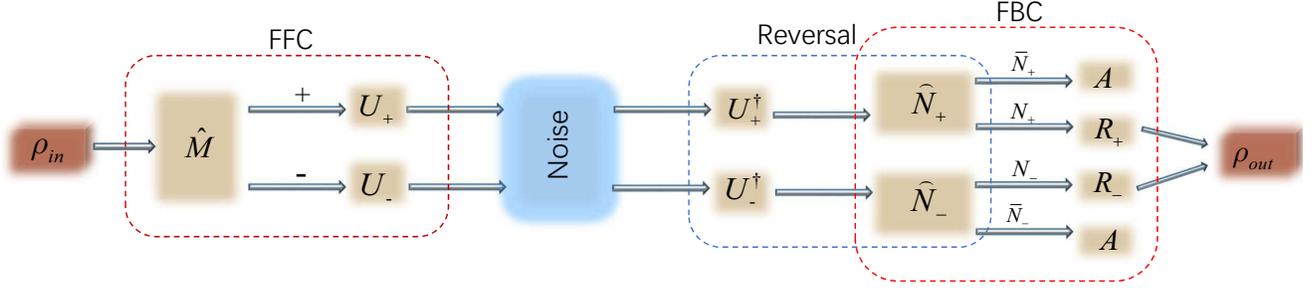}
  \caption{\label{fig:epsart} Schematica of the QCC scheme.}
\end{figure*}


$\mathbf{FFC.}$  Before the noise, Alice firstly makes a preweak measurement $M$ which has two outcomes, i.e., $\pm$.

Then Alice applies a feedforward unitary $U_{\pm}$ according to the measurement result $\pm$. That is, if the result is +, she does the unitary $U_+$; otherwise, she does the unitary $U_-$.

$\mathbf{The \\\ Reversal.}$ After the noise, Bob receives the qubit and makes the reversal of the feedforward control process Alice has done. Concretely, if the measurement result is 1, Bob will firstly apply the reversed unitary $U_+^{\dagger}$ and then a postweak measurement $\hat{N}_+=\{N_+, \bar{N}_+\}$; otherwise, when the result is $-$, he firstly apply the unitary $U_-^{\dagger}$ and the a postweak measurement $\hat{N}_-=\{N_-, \bar{N}_-\}$.

$\mathbf{FBC.}$ Based on different postweak measurement results, Bob then make corresponding corrections. That is, if the result is $\hat{N}_{\pm}$, he does a feedback unitary $R_{\pm}$. Otherwise, he does the action ``$A$'' which is short for abandon.

Our main task is to find appropriate control operators to improve the protection. The design of the measurements and unitaries should satisfy the rules as follows.

(I) The measurements therein should be weak, in order to balance the information gain and back-action. Also these measurements should be strength dependent. So that one can adjust the measurements to satisfy what needs.

(II) The state of the qubit before entering the noise should be close to the immune state of the noise, i.e., the state that won't be affected by the noise. In this paper, the immune state is unique, i.e. $|0\rangle$.

(III) The output state after the FFC procedure should be close to the original state in the sense that the measurement operators $M_+N_+$ and $M_-N_-$ are close to $I$. Thus we call the postweak measurement the reversal of the preweak measurement.

(IV) The FBC correction should correct the error as perfect as possible.

Under requirements above, we design the measurements in a family as follws
\begin{equation}
\begin{array}{l}
   {{M}_{+}}=\sqrt{p}\left| {{V}_{+}} \right\rangle \left\langle  {{V}_{+}} \right|+\sqrt{1-p}\left| {{V}_{-}} \right\rangle \left\langle  {{V}_{-}} \right|,  \\
   {{M}_{-}}=\sqrt{1-p}\left| {{V}_{+}} \right\rangle \left\langle  {{V}_{+}} \right|+\sqrt{p}\left| {{V}_{-}} \right\rangle \left\langle  {{V}_{-}} \right|,  \\
 \end{array}
\end{equation}

where
\begin{equation}
 \begin{array}{l}
\left| {{V}_{+}} \right\rangle =\cos \frac{\alpha +\pi/2}{2}\left| + \right\rangle +e^{i\phi}\sin \frac{\alpha +\pi/2}{2}\left| - \right\rangle,\\
\left| {{V}_{-}} \right\rangle =\sin \frac{\alpha +\pi/2}{2}\left| + \right\rangle - e^{i\phi}\cos \frac{\alpha +\pi/2}{2}\left| - \right\rangle.
 \end{array}
\end{equation}

Comparing to the preweak measurement used in Ref.[], i.e., $M_1=\sqrt{p}|0\rangle\langle0|+\sqrt{1-p}|1\rangle\langle1|$, $M_1=\sqrt{1-p}|0\rangle\langle0|+\sqrt{p}|1\rangle\langle1|$, the measurement above is formed in a larger family by introducing a parameter $\alpha$ encoded in a general basis $\{|V_{+}\rangle,|V_{-}\rangle\}$ other than the logical basis $\{|0\rangle,|1\rangle\}$. A parameter $\alpha$ is introduced in the measurement. The parameter $p$ therein ranges from 0 to 1/2, denote the strength of measurement. Two extreme points describe two trivial measures. When $p=0$, it's a projective measurement. When $p=1/2$, the measurement is $I/2$, doing nothing.

We choose the feedforward unitary that transforms the state to the plane which is less affected by the noise as follows,
\begin{equation}
U_{+}=\left| 0 \right\rangle \left\langle  {{V}_{+}} \right|+\left| 1 \right\rangle \left\langle  {{V}_{-}} \right|,\\
U_{-}=\left| 1 \right\rangle \left\langle  {{V}_{+}} \right|+\left| 0 \right\rangle \left\langle  {{V}_{-}} \right|.
\end{equation}

The postweak measurement with strengthes has the formation of
\begin{equation}
\begin{array}{l}
   \left\{
      \begin{array}{l}
   N_+=\sqrt{1-p_1}|V_+\rangle\langle V_+|+|V_-\rangle\langle V_-|,\\
   \bar{N}_+=\sqrt{p_1}|V_+\rangle\langle V_+|
   \end{array}
   \right.\\
   \left\{
      \begin{array}{l}
   N_-=|V_+\rangle\langle V_+|+\sqrt{1-p_2}|V_-\rangle\langle V_-|,\\
   \bar{N}_-=\sqrt{p_1}|V_-\rangle\langle V_-|
   \end{array}
   \right.
   \end{array}
\end{equation}

After the FFC procedure, the output state will still stay in the plane expanded by the original two states, however slightly deviated from the direction of the original state in the Bloch picture. According to different measurement result, the direction are opposite. Thus our feedback unitary should be a rotation whose axis is orthogonal to the plane expanded by the original state.
\begin{equation}
    R_{\hat{\eta}}(\gamma_{\pm})=\text{cos}(\frac{\gamma_{\pm}}{2})\sigma_0-i\text{sin}(\frac{\gamma_{\pm}}{2})(\text{cos}\phi \sigma_2-\text{sin}\phi \sigma_3),
\end{equation}
where $\{\sigma_i\}_{i=0}^{3}$ is the Pauli basis. $R_{\pm}\equiv R_{\hat{\eta}}(\gamma_{{\pm}})$ are actually rotations about the Bloch vector $\hat{\eta}$ with an angle $\gamma_{\pm}$, where $\hat{\eta}=(0,\text{cos}\phi,-\text{sin}\phi)$ is the Bloch vector of the state $(|+\rangle+ie^{i\phi}|-\rangle)/\sqrt{2}$.

From the view of trajectory, the output state for input state $|\psi_{\pm}\rangle$ will be the mixture of states from different paths,
\begin{equation}
       \rho_{out}^{\pm}= \frac{\sum_{i,j} |\psi_{\pm}^{i,j}\rangle\langle\psi_{\pm}^{i,j}|}{\text{Tr}[\sum_{i,j} |\psi_{\pm}^{i,j}\rangle\langle\psi_{\pm}^{i,j}|]}
\end{equation}
with corresponding success probability
\begin{equation}
    g_{\pm}=\text{Tr}[\sum_{i,j} |\psi_{\pm}^{i,j}\rangle\langle\psi_{\pm}^{i,j}|].
\end{equation}
Here, we have
\begin{equation}
    \begin{array}{lll}
        \psi^{i,j}_{\pm} &=& R_i\cdot N_i \cdot U_i^{\dagger}\cdot E_j\cdot U_i\cdot M_i \cdot \psi_{\pm}.   \\
  \end{array}
\end{equation}

From direct calculation, the average fidelity and average success probability respectively take the form
\begin{equation}
    \begin{array}{ll}
      F&=s_{+}f_{+}+s_{-}f_{-},\\
       &=s_{+}\langle\psi_{\pm}|\rho_{out}^{\pm}|\psi_{\pm}\rangle
    \end{array}
\end{equation}
and
\begin{equation}
    G=s_{+}g_{+}+s_{+}g_{-}.
\end{equation}
respectively.

\section{The Performance}

We have introduce the control scheme for protecting two qubit states with an arbitrary priori probability and two important measures describing how good the scheme performs. However the scheme is presented with parameters. With different sets of parameters, the states will be protected in different ways. Our main task is to figure out how these parameters affect the scheme and how well the generalized scheme can improve the performance of protection.

In this section, we'll show the performance in two aspects. Firstly, we track the state evolution of an initial state under a control scheme, to . We  orfirst consider the general case, i.e., arbitrary prior probability. Then we outline an special case where equal prior probability are given.

\subsection{State evolution}
To make it clear, we track the state evolution of the qubit when the primary state is $|\psi\rangle\equiv|\psi_+\rangle$ and take preweak measurement result $+$ as an example. The process is as follows.

(1): The unnormalized state after getting the preweak measurement $+$  will be
\begin{equation}
    |\tilde\psi_1\rangle=\sqrt{p}\cos \frac{\alpha }{2}\left| {{V}_{+}} \right\rangle +\sqrt{1-p}\sin \frac{\alpha }{2}\left| {{V}_{-}} \right\rangle.
\end{equation}
Here and thereafter, we use $|\tilde{n}\rangle$ to denote an unnormalized form of the state $|n\rangle$.
$$|\tilde{n}\rangle\equiv |n\rangle \langle\tilde{n}|\tilde{n}\rangle.$$

(2): The corresponding feedforward unitary $U_+$ transforms the state to the plane that is less polluted by the noise
\begin{equation}
  |\psi_2\rangle=\sqrt{p}\cos \frac{\alpha }{2}|0\rangle +\sqrt{1-p}\sin \frac{\alpha }{2}|1\rangle.
\end{equation}

(3): The state after the noise can be seen as the combination of two state as follows,
\begin{equation}
    \left\{\begin{aligned}
            |\psi_{3,1}\rangle &= \sqrt{p}\cos \frac{\alpha }{2}|0\rangle +\sqrt{(1-r)(1-p)}\sin \frac{\alpha }{2}|1\rangle,  \\
            |\psi_{3,2}\rangle &= \sqrt{r(1-p)}\sin \frac{\alpha }{2}|0\rangle.  \\
    \end{aligned} \right.
\end{equation}

(4): Trough the reversed feedforward unitary $U_+^{\dagger}$ and  getting a $+$ after the postweak measurement $\{N_+,\bar{N}_+\}$, the state of the qubit will be
\begin{equation}
    \left\{\begin{aligned}
            |\psi_{4,1}\rangle &= \sqrt{p(1-p_1)}\cos \frac{\alpha }{2}|V_+\rangle +\sqrt{(1-r)(1-p)}\sin \frac{\alpha }{2}|V_-\rangle,  \\
            |\psi_{4,2}\rangle &= \sqrt{r(1-p)(1-p_1)}\sin \frac{\alpha }{2}|V_+\rangle. \\
    \end{aligned} \right.
\end{equation}

(5): Finally, Bob applies a "Rotation" $R_1$, which transforms the state into
\begin{equation}
    \left\{\begin{aligned}
            |\psi_{5,1}\rangle &=R_+|\psi_{4,1}\rangle,  \\
            |\psi_{5,2}\rangle &=R_+|\psi_{4,1}\rangle. \\
    \end{aligned} \right.
\end{equation}

\subsection{Arbitrary prior probability}
In this section, we discuss the performance of the presented scheme mainly in two aspects: definite control and indefinite control. In the former one, the average success probability of the scheme is unit and thus for arbitrary input state it can definitely be protected. Inversely, when the success probability is not limited to unit, there is a probability that the scheme will claim to fail for protecting. A definite protection scheme shows the limitation of fidelity when unit success probability is required. In other words, it shows the best protection without any abundant resource consumption. However, we also wonder how well can this scheme perform when resource consumption is not restricted, an indefinite protection scheme. It will be helpful for understanding the overall performance of a state protection scheme and maybe applicable for special cases where fidelity is considered for available scope of success probability. We define the optimal scheme for an indefinite protection scheme as a scheme where parameters achieves the highest fidelity (success probability) when fixed success probability (fidelity) is given. The corresponding parameters are called an optimal indefinite protection parameters set.

According to the expressions of total success probability and total fidelity in Eqs.(\ref{g}) and (\ref{f}), respectively, we find the optimal indefinite protection scheme or parameters instead, we can follow the procedure of the definite protection as discussed above by first settling down the conditions of $p_1$ and $p_2$, then substituting these conditions to the primary interested measure (total fidelity or success probability) to find the optimal $\gamma_{+}$ and $\gamma_{-}$, and finally find the optimal $p$. However, the analytical expressions for these optimal parameters are too complicated, so we choose the numerical way.

To quantify the performance of our indefinite protection scheme, we define a measure of its improvement over the former scheme as below
\begin{equation}
  \begin{aligned}
  \Delta^{dif}_g=f^{opt}_g-\hat{f}^{opt}_g.
  \end{aligned}
\end{equation}
Here $f^{opt}_g$ and $f^{opt}_g$ means the optimal fidelity of the presented scheme and the former one with given $g$.

For every set of parameters $\{ p, p_{1}, p_{2}, \gamma_{+}, \gamma_{-}\}$, there is a unique fidelity and success probability. By searching over all legal ranges of related parameters, we can get a fidelity-success probability phase where all possible values are included. Obviously, the boundary of the phase correspond to the optimal protection schemes. That is to say, every point on the boundary is an optimal protection scheme.

When the measurements are encoded in the logical basis and the feedback rotations are set to be trivial, the quantum composite control scheme will reduce to the pure feedforward control scheme, which has been shown to be capable of archiving unite fidelity when regardless of the success probability. It is straightforward to see that this proposal can also achieve unit fidelity when ranging the success probability from 0 to 1.

To evaluate the advantage of this proposal, we first consider a definite control. We plot figure 2, the improvement of fidelity, the fidelity of this proposal and the optimal measurement angle $\alpha$ for different noise strgenth $r=0.5$ and 1. Results show that the maximum fidelity improvement is about 0.06, occuring at point $(s_+,\theta)=(0,\pi/4)$.

One might think that the best measurement basis is the Helstrom basis. To clarify our proposal to this intuition, we plot figure 3. It's easy to find there is a gap between our proposal and the Helstrom one.

When success probability is not limited to unit, a quantum composite control scheme can in principlly achieve unit fidelity. However, from the view of practical implementation, we would like to pursue higher success probability with fixed fidelity. In figure 4, we plot the relation between fidelity and success probability for this proposal and the one in 2017. Results show that our scheme has an apparent improvement in success probability.

\begin{figure*}[htbp]
\centering
  \includegraphics[width=0.8\paperwidth]{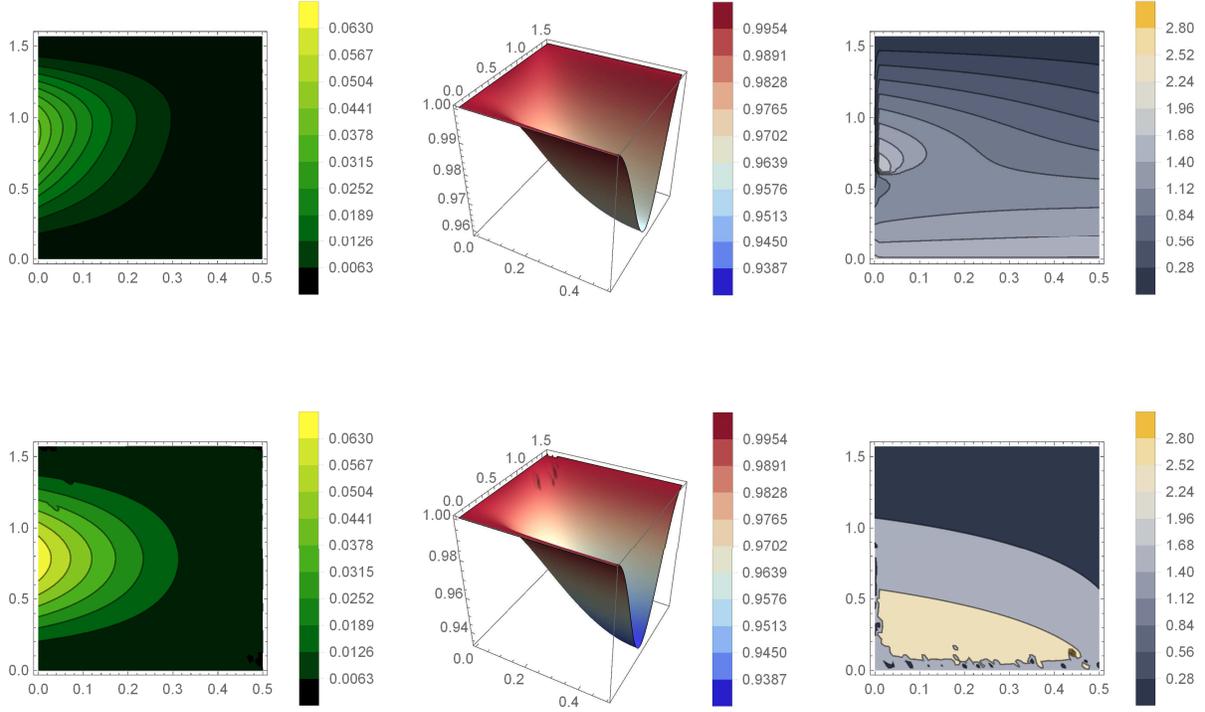}
  \caption{\label{fig:epsart} Fidelity improvement v.s. the initial angle $\theta$ and noise strength $r$.}
\end{figure*}
\label{fig2}

\begin{figure}[htbp]
\centering
  \includegraphics[width=0.4\paperwidth]{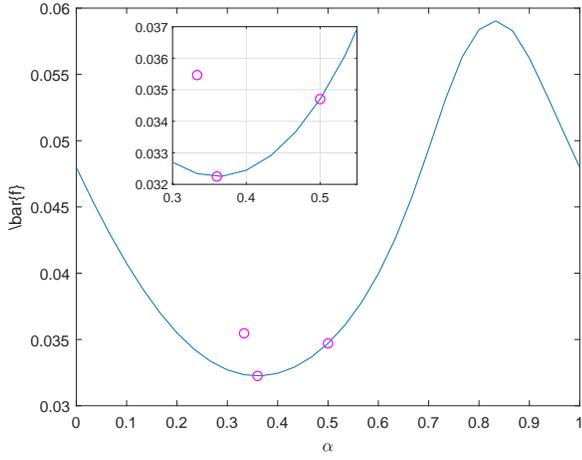}
  \caption{\label{fig:epsart} Curve for infidelity vs. $\alpha$ when choosing $s_+=1/3$, $\theta=\pi/6$, $\phi=0$ and $r=0.8$ of this scheme and points for the optimal GQCC scheme, H scheme and QCC scheme, respectively. }
\end{figure}
\label{fig3}

\begin{figure}[H]
\centering
\includegraphics[width=0.4\paperwidth]{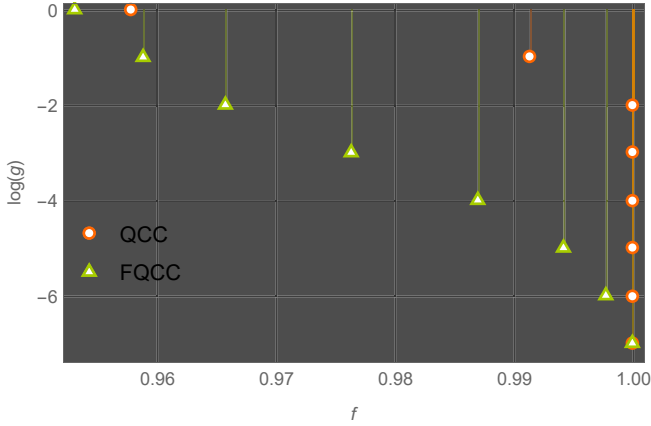}
\caption{Fidelity VS. Log [success probability] when $\theta=\pi/3$ $r$=0.9 and $s_+$=1/3.
Here dot and triangle for the scheme without and with classical side information, respectively.
}
\label{fig:4}
\end{figure}

\subsection{Equal prior probability}

Notice that if we take the strengths of both postweak measurement as
\begin{equation}\label{cond1}
  1-p_{1}=1-p_{2}=1.
\end{equation}
the success probability of the presented scheme will be unit. This is the case which we call a definite protection scheme, where no measurement results are abandoned.

For protecting the initial state $|\psi_{+}\rangle$, the fidelity of this definite protection scheme will be
 \begin{equation}
  \begin{aligned}
  & f_{+}^{'}\equiv f^{'}(a,b)\\
  &=\left| {{r}_{1}}{{a}^{2}}\sqrt{p}-{{r}_{2}}ab\sqrt{(1-r)(1-p)}+{{r}_{2}}ab\sqrt{p}+{{r}_{1}}{{b}^{2}}\sqrt{(1-r)(1-p)} \right|\hat{\ }2 \\
 &+\left| {{s}_{1}}{{a}^{2}}\sqrt{p}-{{s}_{2}}ab\sqrt{(1-r)(1-p)}+{{s}_{2}}ab\sqrt{p}+{{s}_{1}}{{b}^{2}}\sqrt{(1-r)(1-p)} \right|\hat{\ }2 \\
 & +{{b}^{2}}r(1-p){{({{r}_{1}}a+{{r}_{2}}b)}^{2}}+{{b}^{2}}r(1-p){{({{s}_{1}}a+{{s}_{2}}b)}^{2}} \\
  \end{aligned}
 \end{equation}

Replacing $|\psi_{-}\rangle$ with $|\psi_{+}\rangle$ in the presented scheme, we have the fidelity of $|\psi_{-}\rangle$ being $f_{-}^{'}=f^{'}(a^{'}, b^{'})$ with $a^{'}=\cos\frac{\alpha+2\theta}{2}$ and $b^{'}=\sin\frac{\alpha+2\theta}{2}$. To sum up, the total final fidelity of the definite protection scheme for the states of interest becomes
 \begin{equation}
 f^{'}=s_{+}f^{'}(a,b)+s_{-}f^{'}(a^{'},b^{'}).
 \end{equation}
This fidelity is a function of the parameter for priori probability $s_{+}$, the angle of the initial states $\theta$, the amount of the noise $r$, the preweak measurement strength $p$, the feedback rotation angles $\gamma_{+}$ and $\gamma_{-}$. It is interesting to find that the fidelity is independent of the phase of the initial states $\phi$.

When $s_{+}=s_{-}$, $ f^{'}$ reduces to
\begin{widetext}
 \begin{equation}
 f^{'}|_{\pi_{+}=\pi_{-}}(r,\alpha,p,\gamma)
 =\sin\gamma(p+r-pr-\frac{1}{2})\sin\alpha+\cos\gamma[\sin^2\alpha(\sqrt{(p(1-p)(1-r))}-\frac{1}{2})-\cos^2\alpha(1-p)r+\frac{1}{2}]+\frac{1}{2}.
  \end{equation}
\end{widetext}
In this case, the optimal rotation angle $\gamma$ can be given by
\begin{widetext}
\begin{equation}\label{cond2}
 \begin{aligned}
 \gamma_{opt}
 =\arctan\frac{\sin\alpha(p+r-pr-\frac{1}{2})}{\sin^2\alpha(\sqrt{(p(1-p)(1-r))-\frac{1}{2})}-\cos^2\alpha(1-p)r+\frac{1}{2}}
 \end{aligned}
\end{equation}
\end{widetext}

Substituting this optimal $\gamma_{opt}$ into the reduced total fidelity, we have
\begin{widetext}
 \begin{equation}
 f^{'}|_{\pi_{+}=\pi_{-}}(r,\alpha,p)
 =\sqrt{(p+r-pr-\frac{1}{2})^2\sin^2\alpha+[\sin^2\alpha(\sqrt{(p(1-p)(1-r))}-\frac{1}{2})-\cos^2\alpha(1-p)r+\frac{1}{2}]^2}+\frac{1}{2}.
  \end{equation}
\end{widetext}

By straightforward calculations, the optimal $p$ of this definite protection scheme is the real positive root of the quartic equation in parameter $p$ as below
\begin{widetext}
 \begin{equation}\label{cond3}
 (y_{1}^{2}+y_{3}^2)p^{4}+(2y_3y_4+2y_1y_2-y_1^2)p^3+(y_4^2+2y_3y_5-2y_1y_2+y_2^2)p^2+(2y_4y_5-y_2^2)p+y_5^2=0.
 \end{equation}
\end{widetext}
where $y_1=x_1^2-x_3^2+x_4^2, y_2=x_1x_2+x_4x_5+x_3^2/2, y_3=2x_3x_4, y4=x_3x5-3x_3x_4/2, y_5=-x_3x_5/2$ with $x_1=(1-r)\sin\alpha, x_2=(r-1/2)\sin\alpha, x_3=\sqrt{1-r}\sin^2\alpha, x_4=r\cos^2\alpha, x_5=(1/2-r)\cos^2\alpha$.

We call the set of parameters $(p,\alpha,\gamma)$ satisfying conditions in Eqs.(\ref{cond1},\ref{cond2},\ref{cond3}) is an optimal definite protection for initial states and noise with parameters $\theta$, $\phi$ and $r$, and the corresponding scheme an optimal definite protection scheme.

The difference of It is interesting to find that this scheme is better than that of the composite control scheme [] even when equal a priori probability is considered.

With the analytical expression we obtained above, we show this interesting fact by figuring the difference of fidelity between this optimal definite protection scheme and that in a composite control one with the fixed unit success probability in FIG. 2. It can be seen that, advantages are apparent for all $\theta$ and $\phi$ with special $r$. There is a trend that, the heavier the noise is, the more advantages there are. When it is complete decoherence, the fidelity improves a generous number about 0.5.

\begin{figure}[htbp]
\centering
  \includegraphics[width=0.4\paperwidth]{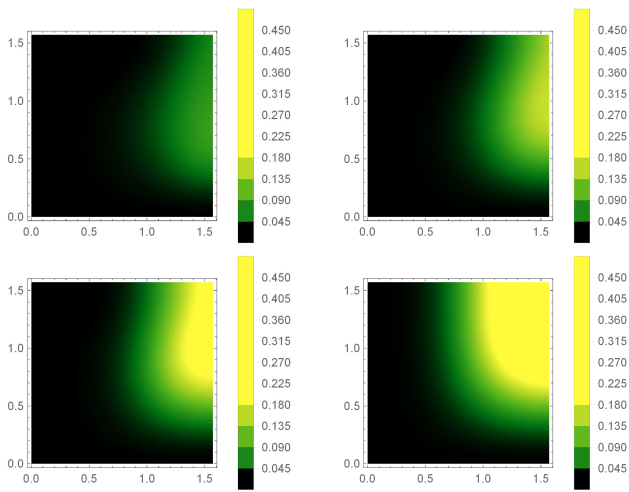}
  \caption{\label{fig:epsart} Fidelity improvement v.s. the initial angle $\theta$ and noise strength $r$.}
\end{figure}
\label{fig6}

\section{Conclusion}
In this paper, we introduce a side information -driven quantum composite control scheme. By consideration of all given side information, we design the measurements and unitaries therein and improve the performance of protecting the state of a qubit. The performance has been discussed for both analysis and numerical ways by optimization over parameters in initial states, the noise and the scheme. The advantages lies in the following aspects. Firstly, it works for a wider range of states and generalizes our last work. The states we protect here are more general and complements the scale for state protection and thus extends the application of quantum composite control. Secondly, we exploit measurements and corrections by mixing the information of not only the initial angle, but also the initial phase and priori probabilities, which leaves room to improve the protection performance. Analysis shows that this proposal outperforms in two aspects. On the one hand, this proposal improves the fidelity when fix a success probability for all initial states and noises. On the other hand, it successes with the highest probability when fixing a fidelity for all initial states and noises. Even when equal a priori probability, this proposal shows improvements over the previous quantum composite control scheme. At last, we expect implementation of this generalized scheme and its application in various techniques, such as quantum state discrimination, quantum storage and so on.

\begin{acknowledgments}
 This work is supported by NSFC (Grant Nos. 61572081, 61672110, 61671082).
\end{acknowledgments}

\nocite{*}

\bibliography{apssamp}

\end{document}